\documentclass[10pt]{article}

\usepackage{amsmath}
\usepackage{amssymb}

\usepackage{graphicx}

\usepackage{cite}

\usepackage{color} 

\usepackage[labelfont=bf,labelsep=period,justification=raggedright]{caption}

\makeatletter
\renewcommand{\@biblabel}[1]{\quad#1.}
\makeatother

\date{}

\begin{document}

\begin{flushleft}
{\Large
\textbf{A DATA DRIVEN NETWORK APPROACH TO RANK COUNTRIES PRODUCTION DIVERSITY AND FOOD SPECIALIZATION}
}
\\
Tu Chengyi$^{1}$, 
Joel Carr$^{2}$, 
Samir Suweis$^{1,\ast}$
\\
\bf{1} Department of Physics and Astronomy, University of Padova, Padova, 35131, Italy
\\
\bf{2} Department of Environmental Sciences, University of Virginia, Charlottesville, 22904-4123, USA
\\

$\ast$ E-mail: suweis@pd.infn.it
\end{flushleft}


\section*{Abstract}
The easy access to large data sets has allowed for leveraging methodology in network physics and complexity science to disentangle patterns and processes directly from the data, leading to key insights in the behavior of systems. Here we use to country specific food production data to study binary and weighted topological properties of the bipartite country-food production matrix. This country-food production matrix can be: 1) transformed into overlap matrices which embed information regarding shared production of products among countries, and or shared countries for individual products, 2) identify subsets of countries which produce similar commodities or subsets of commodities shared by a given country allowing for visualization of correlations in large networks, and 3) used to rank country's fitness (the ability to produce a diverse array of products weighted on the type of food commodities) and food specialization (quantified on the number of countries producing that food product weighted on their fitness). Our results show that, on average, countries with high fitness producing highly specialized food commodities also produce low specialization goods, while nations with low fitness producing a small basket of diverse food products, typically produce low specialized food commodities.

\section*{Introduction}

By providing powerful theoretical tools and incentivizing innovative steps towards the comprehension and synthesis of broad empirical facts from increasingly available datasets, networks physics and complexity science are actively contributing to our understanding of ecological \cite{montoya2006ecological, dominguez2015ranking}, environmental \cite{o2014globalization, seaquist2014architecture, suweis2013water, suweis2015resilience}, economic \cite{caldarelli2012network, cristelli2013measuring, cristelli2015heterogeneous, tacchella2012new} and social \cite{liljeros2001web, suweis2014early} processes.
In ecology, scientists have shown that ubiquitous emergent patterns found in distinctly different ecosystems can be described by a few basic ecological processes (e.g. birth, death, migration, adaptation and niche differentiation) which in turn can be incorporated into simple conceptual models \cite{williams2000simple, bastolla2009architecture, martin2015eluding, azaele2015statistical}.
In a similar manner, patterns of human mobility which incorporate complex details of social and individual behavior can be readily described by simpler statistical models \cite{newman2001structure, simini2012universal}.
As another example, it has been shown that only few economic or environmental variables are necessary to reconstruct many properties of corresponding trade networks and quantify the statistics of the relative fluxes of goods (measured in \$) or of water (using virtual water \cite{konar2011water}) among nations \cite{garlaschelli2004fitness, suweis2011structure}.

These examples all demonstrate how quantitative data driven approaches allow for disentangling the key drivers that produce a given result.
An interesting example illustrating the above claim, is a series of recent articles in which, based on economic data, the authors were able to measure the intangible "capabilities" of countries and define the economic product "complexity" \cite{hidalgo2009building, caldarelli2012network, cristelli2013measuring, cristelli2015heterogeneous, tacchella2012new}.
Hidalgo and Hausmann \cite{hidalgo2009building} have been the first to introduce the idea that the diversity of products and a nation's capabilities can be inferred directly from the structure of the country-product matrix.
In their view, ubiquitous products require little sophistication and thus can be produced by most countries and diversified countries possess many from capabilities allowing production of sophisticated (more complex) commodities.
From this perspective, the most diversified nations are expected to be the high ranked in terms of global economic competition (or fitness).
Similarly, ubiquitous products are more likely correspond to products with low sophistication.
To quantify these concepts, Hidalgo and Hausmann \cite{hidalgo2009building} introduced the so-called "reflections method" through which they obtain country economic competitiveness and production complexity \cite{hidalgo2009building}.
Unfortunately, this method suffers from conceptual and practical problems as demonstrated in Cristelli et al. \cite{cristelli2013measuring}.
To overcome these problems, recent efforts \cite{caldarelli2012network, cristelli2013measuring, tacchella2012new} have defined a new method to determine the relative strength of countries and export products.
This method is based on the introduction of coupled non-linear maps between country fitness and product complexity (see below for a more detailed definition), and are characterized by a fixed point.
By exploiting the information contained in a matrix comprised of the detailed export of each country and iteratively combining measures on the matrix rows and columns, the authors were able to extract the relation between the export basket of a given country and how it relates to that country's economic competitiveness and product complexity.
This relationship was expected to exist based on an established economic idea known as the Ricardo hypothesis \cite{ricardo1891principles}.
The Ricardo hypothesis suggests that: highly economic competitive nations should export only products comprised of a high level of sophistication (e.g. airplanes), while poor countries should export only the products which require no special capabilities (e.g. hammers).
Contrary to Ricardo hypothesis, the above studies found that most developed countries export almost all products from simple (including those not requiring any level of technology) to complex, while poorly developed countries were able to only produce and subsequently export only relatively simple products (e.g. products which require little technological sophistication).

Inspired by these works, herein we examine the relationship between countries and the food products they produce.
As there clearly exist constraints to any country's food production capabilities from climate, land, water and technology, is there an analog of the Ricardo hypothesis when considering food production?
By generalizing the theoretical framework presented in Tacchella et al \cite{tacchella2012new}, we seek to answer the following questions:
1) Are there ubiquitous  or "low specialization" food commodities that are produced by the majority of countries (the analog of the hammers)?
2) On the other hand what are the "high specialization" food products that only a select few countries are capable of producing (an analog to airplanes)?

We also seek to evaluate analog of economic fitness for each nation, but in terms of food production capability.
Can we identify high fitness countries that are able to sell not only several different food products, but food products that are also high specialized (as they can place on the market food commodities that the other nations are not able to produce)?
In contrast then can low fitness countries produce only food commodities of low specialization?
Using the proposed theoretical framework, we set out to generate an unbiased quantification of both specialization for all food products and fitness for all the countries.


\section*{Results}

\subsection*{Country-Food products network properties}

\begin{figure}[!ht]
\centering
\includegraphics[width=0.9\textwidth]{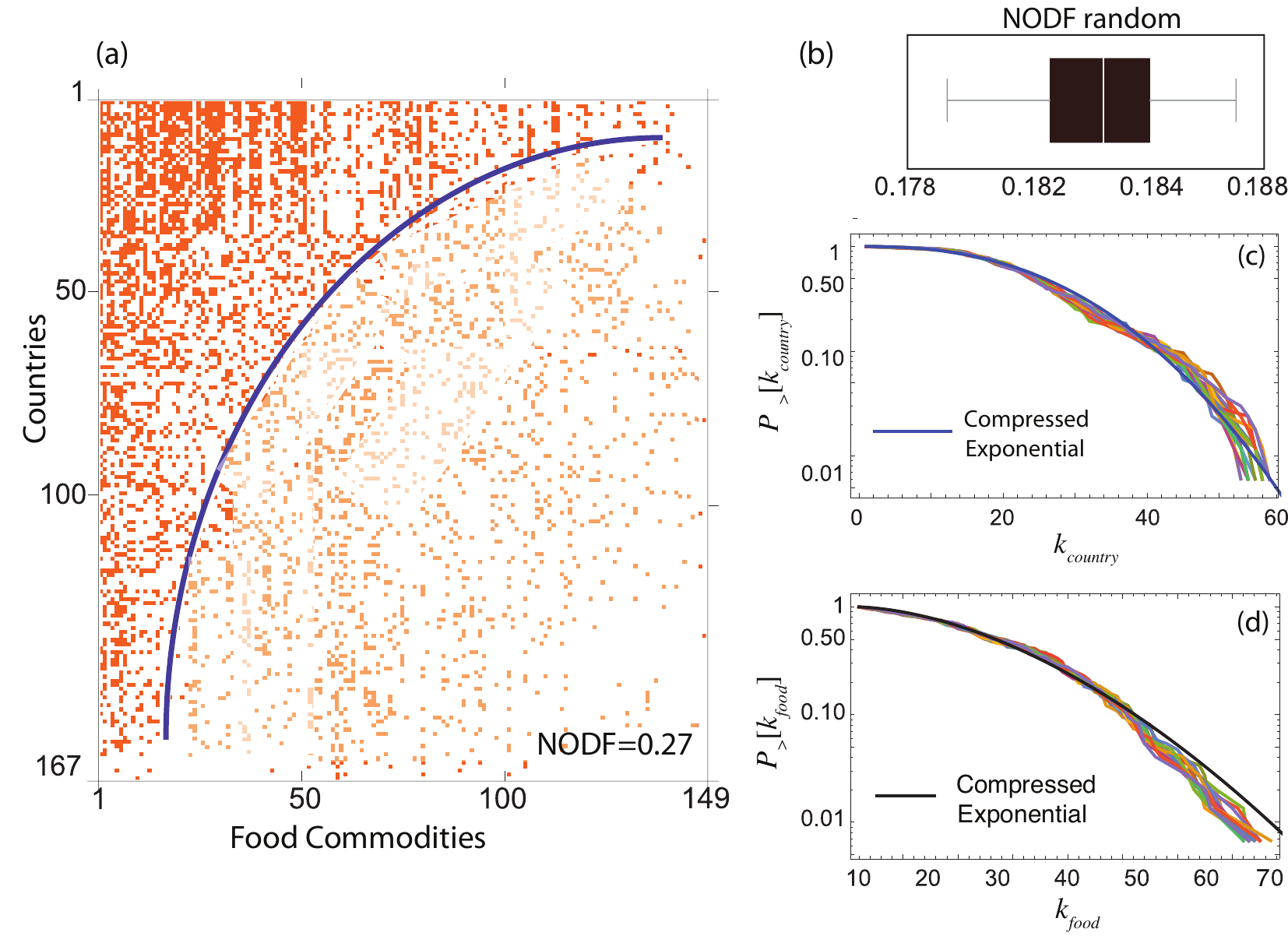}
\caption{(a) The nested structure of the country-food production network. The x-axis and y-axis are the food commodities and countries, respectively. Every element (i,j) in the matrix is one (in orange) if country i produces food commodity j;
(b) Box-whisker plot of nestedness value based on 1000 randomly assembled networks of the same size and connectance of the empirical network. The ends of the whiskers represent the minimum and maximum, whereas the ends of the box are the first and third quartiles and the black bar denotes the median. The plots show that the country-food production networks have nestedness values consistently higher than randomly assembled networks; (c) The cumulative in-degree and (d) out-degree distribution of the country-food production network for the years 1992-2011 (colored lines). The cumulative in-degrees and out-degrees are distributed consistently with compressed exponential distributions (in black), i.e., Weibull distribution with parameters $\alpha = 2.510$, $\beta = 30.000$, $\mu = -0.387$ and with $\alpha = 1.776$, $\beta = 32.265$, $\mu = 0.586$, respectively.}
\label{fig:Figure1NEW}
\end{figure}

In order to obtain a first order understanding of the relationships among countries induced by their food production (and vice-versa), we can study binary and weighted topological properties of the bipartite country-food production matrix (Figure \ref{fig:Figure1NEW} (a)).
This type of analysis is relatively standard \cite{newman2010networks}, and thus herein we only analyze a few properties that are relevant for the main objectives of this work.
Let ${M_{cp}}$ be the adjacency matrix of the country - food production graph.
Then the number of food commodities produced by country $c$ is  $k_c^P = \sum\nolimits_p {{M_{cp}}}$ while the number of countries that produce food commodity $p$ is $k_p^C = \sum\nolimits_c {{M_{cp}}}$.
Thus the number of food commodities produced both by country $i$ and $j$ is $o{}_{ij}^P = \sum\nolimits_p {{M_{ip}}{M_{jp}}}$ ($o$ is known as overlap matrix \cite{suweis2013emergence}).
Analogously, the number of countries that produce both food products $u$ and $w$ is $o{}_{uw}^C = \sum\nolimits_c {{M_{cu}}{M_{cw}}}$.
Nestedness \cite{guimaraes2006improving} is a global network measure that is a function of the above overlap and describes a hierarchy in the interactions in ecological networks (e.g. mutualistic plant-pollinators communities \cite{dominguez2015ranking, suweis2013emergence}).
If the Ricardo hypothesis holds in our case, then low fitness countries produce low specialization food products, while nations with high fitness only produce specialized food commodities.
In this regard the corresponding country-food production graph should not be nested, but rather compartmentalized (i.e., modular \cite{newman2010networks}).
In partial agreement with what is found for industrial and economic products \cite{bustos2012dynamics}, figure \ref{fig:Figure1NEW} (a)-(b) shows that bipartite country-food product networks have nestedness values consistently higher than those found in randomly assembled networks (P-value $<0.001$ ) for all the years analyzed (1992-2011).
Therefore, food commodities produced by countries that are able to produce only a small basket of different food products, are also produced by nations that are characterized by large and diversified food production.
The nested topological structure thus indicates that specialized food commodities (low $k_{food}$) are in general produced by countries with high food diversity (high $k_{country}$), while generalist "food commodities" (low specialization and high $k_{food}$) are also produced by "specialist" countries (low $k_{country}$).
We also found that both $k_{food}$ and $k_{country}$ are distributed consistently with a Weibull distribution (with parameters $\alpha = 2.510$, $\beta = 30.000$, $\mu = -0.387$ and $\alpha = 1.776$, $\beta = 32.265$, $\mu = 0.586$, respectively).
Both $k_{food}$ and $k_{country}$ display a compressed exponential decay in the tail implying a relatively limited heterogeneity in terms of the number of products produced by a single country, and on the number of countries producing the same food product, respectively.

\subsection*{Projection graphs and minimum spanning forest}

\begin{figure}[!ht]
\centering
\includegraphics[width=0.8\textwidth]{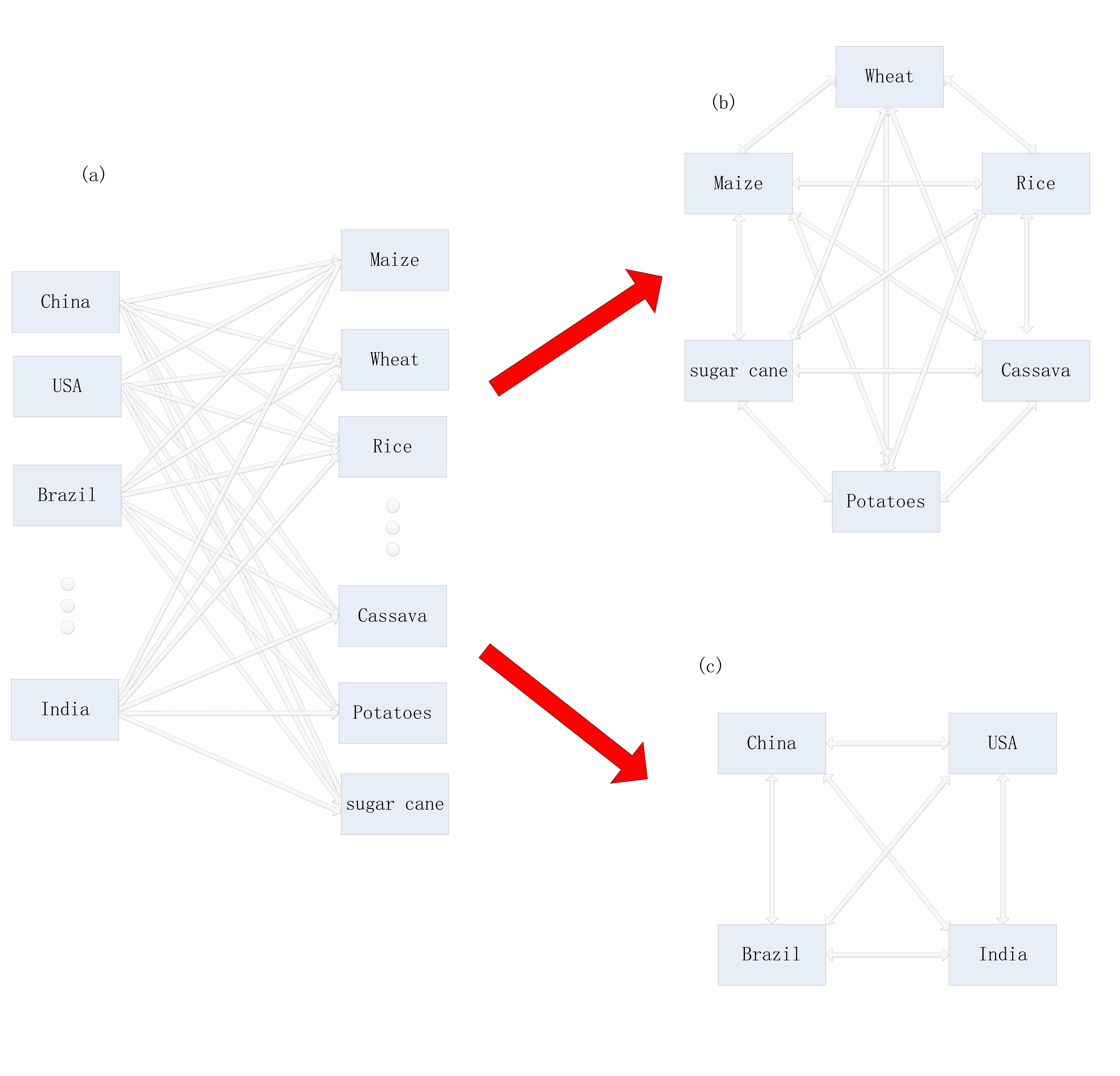}
\caption{A simple example from country-food production network to two projection networks.
(a) bipartite network of country and corresponding production of food commodities;
(b) food-food network where nodes represent food commodities and edges represent food commodities produced by the same country;
(c) country-country network where nodes represent countries and edges represent countries produce the same food commodity.}
\label{fig:Figure2NEW}
\end{figure}

Important information on food production patterns can be also obtained from the weighted country-food production matrix $W$ by building projection graphs (see Methods section and Figure \ref{fig:Figure2NEW} (b) and (c).
In projection graphs, nodes represent countries (or food commodities) and we place a link if two countries produce the same food commodity (or if two food products are produced by the same country).
This connects the various countries (or food products) with a link whose strength is given by the number of products they mutually produce (or the number of countries producing the same given food).
In such a way the information stored in the matrix W is projected into the network of country-country and food-food products as shown in Figure \ref{fig:Figure2NEW}.

By defining a similarity matrix (see Methods section), we can quantify the correlation between any two countries (or food commodities).
Following Caldarelli et al.\cite{caldarelli2012network}, we find the Minimal Spanning Tree (MST) of our projection networks (with the constraint that no edge between two nodes is allowed if they have already been connected to some other node) in order to visualize only the strongest correlations among countries or food products.
This method generates a set of disconnected sub-trees (i.e. a forest) embedded within the MST (details of this algorithm, called Minimal Spanning Forest (MSF), is explained in the Methods section).

\begin{figure}[!ht]
\centering
\includegraphics[width=0.8\textwidth]{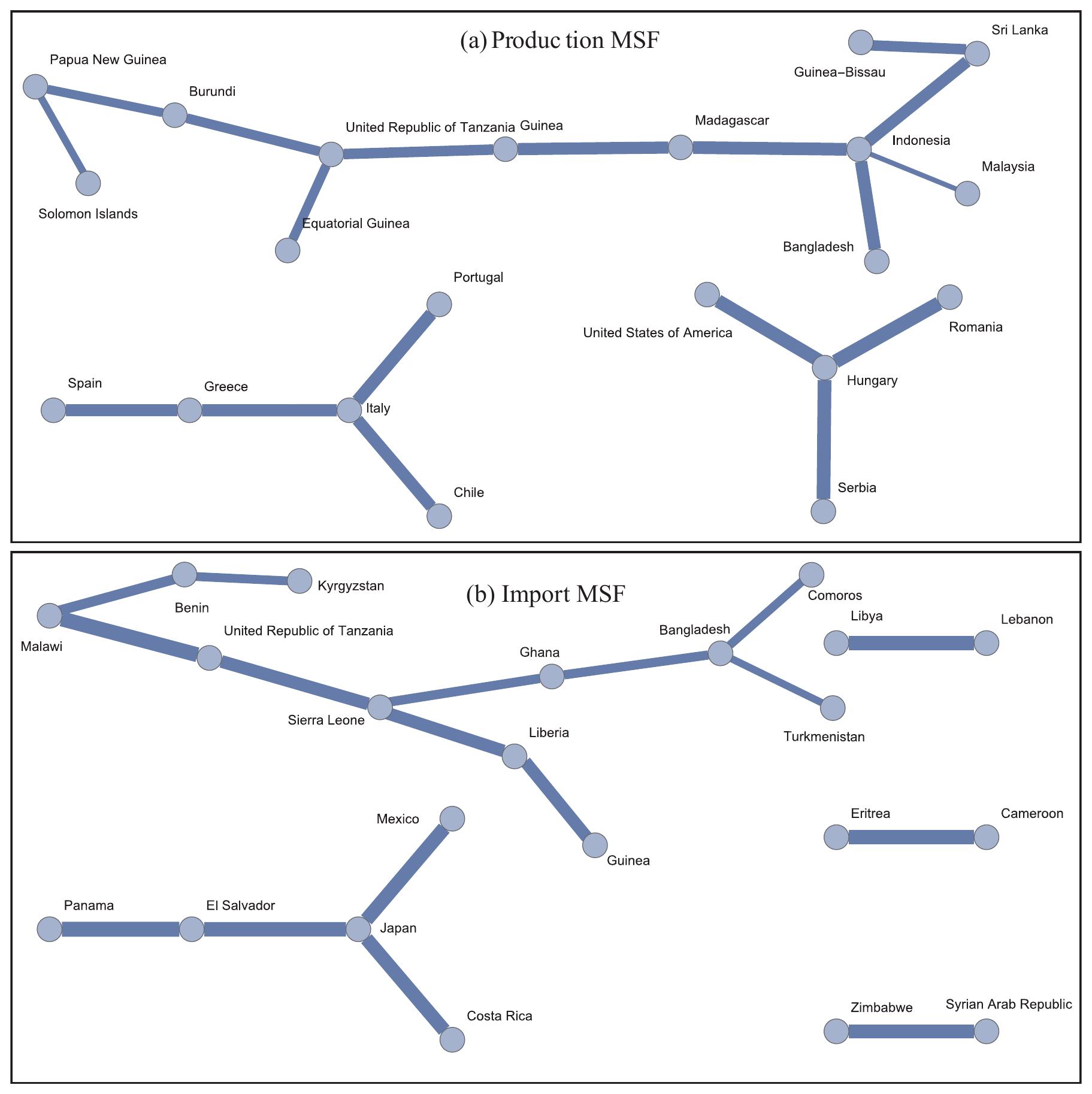}
\caption{(a) Three sub-trees from the MSF of country-food production network. Each sub-tree contains countries that produce very similar products;
(b) Five sub-trees from the MSF of country-food import network. Each sub-tree contains countries that compete in the food market to import the same pool of food commodities. The thickness of each edge represents the similarity strength between the connected countries. }
\label{fig:Figure3NEW}
\end{figure}

\begin{figure}[!ht]
\centering
\includegraphics[width=0.8\textwidth]{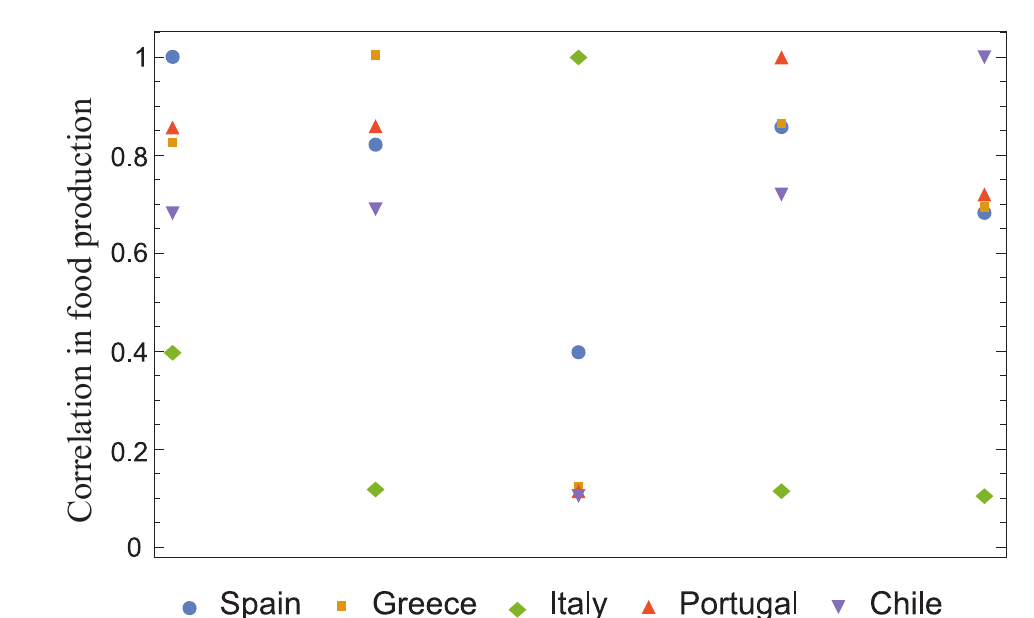}
\caption{Pearson correlation among the food production patterns of the five countries belonging to the second largest sub-tree in Figure \ref{fig:Figure3NEW} (a). We can see that the MSF methods highlight correctly the correlation structure in the projections networks. The Pearson correlation has been calculated using all food commodoties for each pair of countries, i.e. $r = \sum_{j =1}^{N_P} (x_j-\bar{x})(y_j-\bar{y})/N_P$, where $N_P$ is the total number of food commodoties, $x_j$ and $y_j$ are the volumes produced of the commodity by the two countries.}
\label{fig:Figure4NEW}
\end{figure}

In this manner,the MSF naturally splits the network of countries into separate subsets allowing for the visualization of correlations in a large network.
For all years considered in this study (1992-2011), we find many small sub-trees with geographical and/or climatological similarities (see Figure \ref{fig:Figure3NEW}) that lead to significant correlations in the food production and import patterns and only a few large sub-trees(see Figure \ref{fig:Figure4NEW} and Table \ref{tab:table1NEW}).
We emphasize that by construction, the countries comprising each forest are correlated in their production and thus potentially compete in the food market.
In this context, by analyzing all MSFs, we found that developing countries primarily appear to be direct competitors of their geographical neighbors.
However, edges among some countries depend on climatological similarities rather than geographical vicinity (Figure \ref{fig:Figure4NEW}).

\subsection*{Food Specialization and Country Fitness}

We now introduce the iteration process proposed in Tacchella et al. \cite{tacchella2012new} which leverages the structure of the weighted country-food production matrix W.
This method seeks to quantify a country's fitness $F_c$ - a measure of its ability to produce a diversified and specialized food basket - and to evaluate food production specialization $Q_p$ - i.e., food products that are produced only by few high-fitted countries.
The basic idea is to define an iterative process, which couples and estimates $F_c$ and $Q_p$.
Final estimates of $F_c$, $Q_p$ are achieved in the limit of large number of iterations.
This non-linear iterative map can be summarized in the following way.
Firstly, the country fitness $F_c$ is assumed to be proportional to the sum of the different food commodities produced in that nation, weighted by their specialization $Q_p$.
As a first approximation, $Q_p$ is considered to be inversely proportional to the number of countries which produce it, weighted by their inverse fitness (i.e. if a country has a high fitness this should reduce the weight in calculating the specialization of a product, and viceversa \cite{tacchella2012new}).

Iteration between these two calculations converges to estimations of each country's fitness and the food product specialization (see the Method section for the complete algorithm).
The values of $F_c$ and $Q_p$ are quite heterogeneous among nations and products (i.e., after 50 iterations of the non-linear map, they evolve and their distributions follow a fat tail, LogNormal like distribution (see Figure \ref{fig:Figure7NEW}).
Thus the non-linear iterative mapping leads to a unique asymptotic solution (i.e. fixed point) allowing for ranking countries and food commodities based on their respective fitness and specialization.

\begin{figure}[!ht]
\centering
\includegraphics[width=0.8\textwidth]{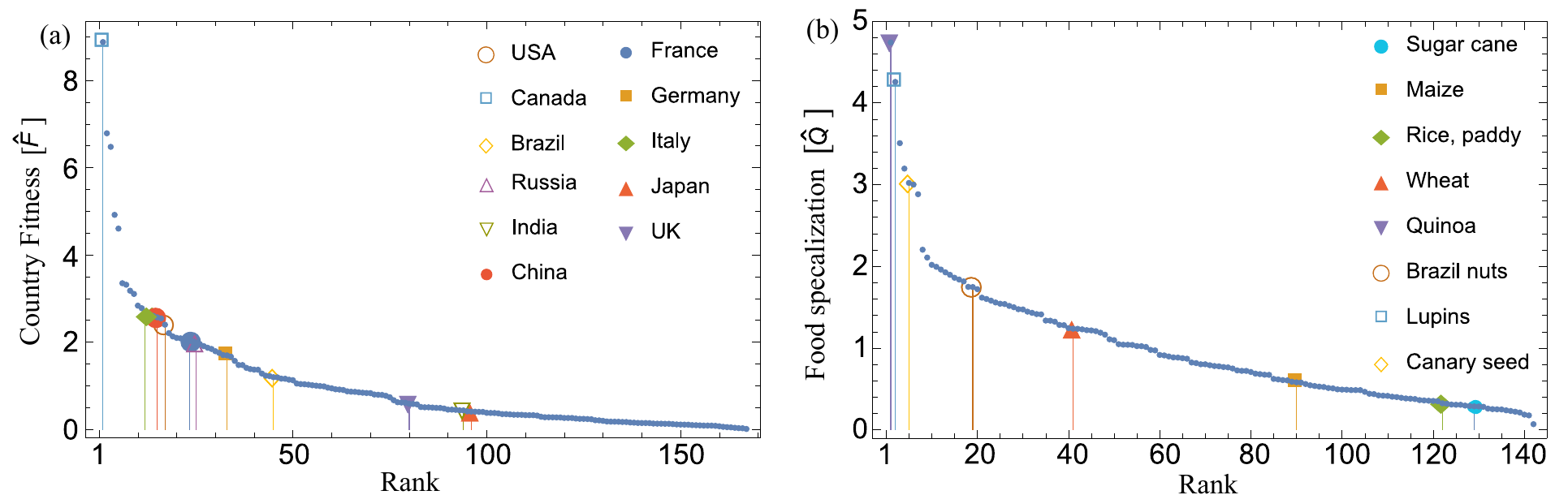}
\caption{(a) The rank of country fitness of some representative countries. In general, the rank of G7 countries are quite high. In BRIC, the rank of China and Russian Federation is high, Brazil is in the middle and India is quite low.
(b) The rank of production complexity of some representative food commodities. On one hand, the production complexity of "Sugar cane", "Maize"", "Rice, paddy" and "Wheat" is low because their yield is very large; On the other hand, the production complexity of "Quinoa", "Brazil nuts, with shell", "Lupins" and "Canary seed" is high because they are rare products.}
\label{fig:Figure5NEW}
\end{figure}

While we can expound on the fitness and specialization of all nations we limit our textual analysis to socioeconomic political groupings such as the G7 (France, Germany, Italy, Japan, United Kingdom, United States of America and Canada) and BRIC (Brazil, Russian Federation, India and China) shown in Figure \ref{fig:Figure5NEW} (a).

As would be expected, country fitness of G7 countries is in general quite high (see Figure \ref{fig:Figure5NEW} and Table \ref{tab:table2NEW}).
However there are important differences: United Kingdom and Japan are always at the bottom of the ranking, while fitness of Canada is always top.
This can be explained by UK and Japan being island nations with territory that is not very extensive; and consequently low variability in climate and geography relative to other nations.
This leads to a relatively low capacity to produce a large quantity of diverse food commodities (in fact they have a below-average weighted degree: 0.41, 0.98  tons/(person*year, respectively).
In contrast Canada is large geographically, stretching from the Atlantic Ocean in the East to the Pacific Ocean in the West, and comprised of diverse climatic regions, from temperate on the West coast of British Columbia to a subarctic climate in the North.
This allows for production that is various and abundant (it has a weighted degree of 2.68 tons/(person*year)).
In the BRIC countries (see Table \ref{tab:table2NEW}), India has high variability in climate and geography, but its rank of country fitness is quite low.
It reflects an insufficient diversified food production with respect to its large population. 
On the other hand, China and Russian Federation also possessing variable climate and geography, are on the top of the rank.  For these two countries, the capacity to produce a large quantity of diverse food commodities is high.
That is to say, for food product the productivity of labour of India is low, while China and Russian Federation is high.

Similarly, Figure \ref{fig:Figure5NEW} (b) presents the rank of production specialization for four crops with large yield ("Sugar cane", "Maize", "Rice, paddy" and "Wheat") and four products with high production specialization ("Quinoa", "Brazil nuts, with shell", "Lupins" and "Canary seed").
The production specialization of the food commodities with very large yield is low (low Q), as many countries produce these products (high $k_{country}$).
On the other hand, the rank of some rare products, such as Quinoa and Lupins, is high because only a few countries could produce them, given the particular climatic and soil condition needed to produce these types of crops.

\section*{Discussion}

This work applied tools and concepts from complex network science to analyze the world food production data.
Based on a country-food production matrix, we analyzed the country-food product network properties and correlation among countries using MSF methods and calculated the country fitness and food production specialization.

Our results show that there exists a hierarchical nested organization in the country-food production network.
The significant nested structure observed in the country-food production graph highlights how there are many "basic" (low specialization) food products that are produced by most of the countries (the analog of "the hammers"), and few complex commodities characterized by particular climatic or geographical conditions (e.g. "the airplanes") that are produced only by few countries. Therefore, also in our case the Ricardo hypothesis does not hold, as the most fitted countries produce almost all products from the common to the specialized one.
Wheat, rice, maize or sugar cane are staple commodities that are the building blocks of our "food" economy whereas, Quinoa, Brazil nuts, and Lupins are produced only by few countries (e.g. Bolivia, Peru) which in turn, although they are not the large food producers (in terms of total volumes), they produce many different food products (high $k_{country}$)  and are also specialized in producing these "complex" commodities. The observed structure reflects the existence in a trade-off between diversification in food production, and intensification of some "optimal" commodity based on climatic-soil-water conditions.

We found that both degree distributions and NODF values are very stable among different years (Figure  \ref{fig:Figure1NEW} (c)-(d)), suggesting that the topological structure of the country-food production network has not changed significantly during the last twenty years.
We leveraged a MSF algorithm in order to highlight the correlation among countries and volumes of food produced resulting in the discovery that countries with high GDP, as for example United States of America, most often belong to small sub-trees.
In general, as the food production basket of these developed countries is large, it is difficult to have a high similarity with all other countries as similarity depends on the relative shared volumes.
Alternatively, countries with relatively small diversity in food production, but with a pool of typical (high volume) food products tend to belong to larger sub-trees.
As an example (Figure \ref{fig:Figure3NEW} (a)) Indonesia, Sri Lanka, Guinea-Bissau, Papua Guinea, Madagascar and Burundi, are very far in term of geographical distance, but in all these countries rice is the main and staple food  product  and as such they all belong to the largest sub-tree.
Similarly, the second largest subtree in composed mainly by countries with a Mediterranean-type of climate and diet.
The consequence is that the MSF detects significant correlation structure in the production patterns of these countries (see Figure \ref{fig:Figure4NEW}).

\begin{table}[!ht]
\centering
\begin{tabular}{|l|l|l|l|}
\hline
  & Maize & Wheat & Rice, paddy\\
\hline
Panama & 0.55961 & 0.24978 & 0.10907 \\
\hline
EI Salvador & 0.55548 & 0.24181 & 0.08627 \\
\hline
Japan & 0.65401 & 0.18581 & 0.00008 \\
\hline
Mexico & 0.43061 & 0.16219 & 0.03848 \\
\hline
Costa Rica & 0.56965 & 0.15158 & 0.04965 \\
\hline
\end{tabular}
\caption{\label{tab:table1NEW}The three largest import food commodities of the countries in the second largest sub-tree in Figure \ref{fig:Figure3NEW} (b). The first column is these five countries and each row is five largest import food commodities for one country.}
\end{table}

Applying the same approach to the country-food import network and building the MSF for imported food volumes (rather than for food production) gives information on the sets of countries that compete in the food market to import from the same pool of food products (see Figure \ref{fig:Figure3NEW} (b)).
Table \ref{tab:table1NEW} shows the three largest import food commodities of the countries in the second largest sub-tree, Panama, El Salvador, Japan, Mexico and Costa Rica with clear similarity in the food commodities that these countries import.
We speculate that if there is a price spike on given food commodity shared within a sub-tree (such as Wheat in Table \ref{tab:table1NEW}), competition members of that sub-tree whom are characterized by large import of same food commodity will dramatically increase.
This has the potential to trigger food crises in poorest nations (those nations with limited ability to compete financially in the import market) of the corresponding sub-tree.

Our methods also provided an estimate quantification of both specialization for all food products and fitness for all the countries.
We note that in our analysis, using data from 1991 to 2011, Quinoa emerges as one of the most specialized food commodity. It is interesting to note that in the October 2015, imports of Quinoa from Europe increased by 40\%: the same growth rate as recorded both in 2013 and in 2014 \cite{Eurostat}. Therefore our analysis allowed us to identify specialized products with high growth potential in the food market.

\begin{figure}[!ht]
\centering
\includegraphics[width=0.8\textwidth]{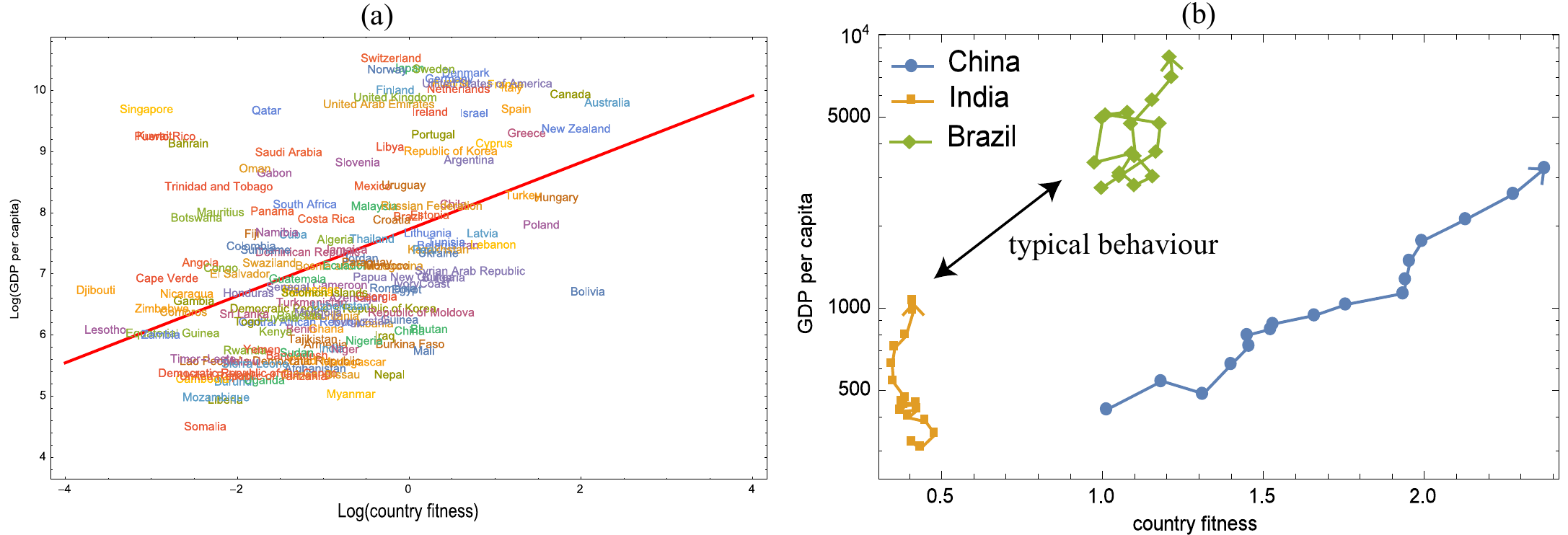}
\caption{(a) The static country fitness - GDP per capita plane of the year 2011. We do not find significant correlation between these two quantities; for example Singapore has high GDP per capita but low fitness. On the other hand Bolivia has high fitness (world leader producer of Quinoa), but low GDP per capita.
(b) The dynamical evolution of country fitness - GDP per capita of China, India and Brazil from 1992 to 2011. As time goes on, the GDP per capita of these countries increase, but the fitness of Brazil and India are constant on average while China has increased.}
\label{fig:Figure6NEW}
\end{figure}

\begin{figure}[!ht]
\centering
\includegraphics[width=0.8\textwidth]{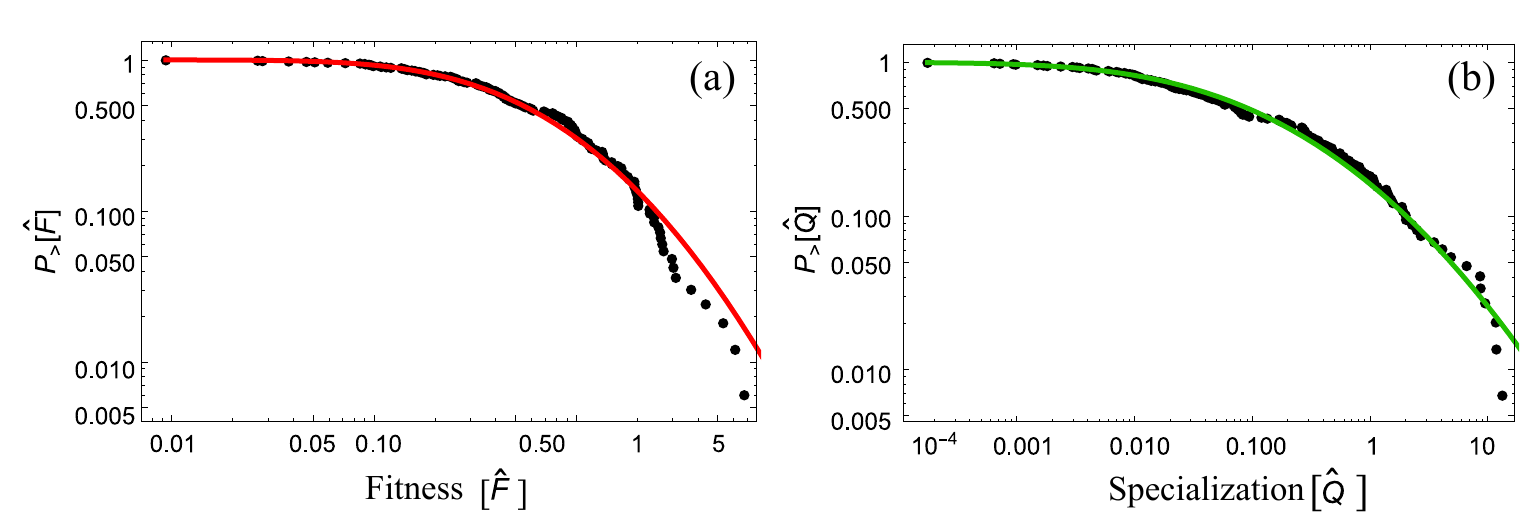}
\caption{The cumulative distribution of (a) country fitness and (b) production specialization. The country fitness and production specialization are from the initial state towards Log-Normal distribution with parameters $\mu=-0.607839$, $\sigma=1.18597$ and $\mu=-2.37283$, $\sigma=2.40229$, respectively.}
\label{fig:Figure7NEW}
\end{figure}

It is interesting to see if there is any correlation between this new network measure (in particular country fitness) and other economic or food security indicators.
A strong correlation would suggest that country fitness is somehow a redundant information with respect to prior well-known indicators such as per-capita Gross Domestic Product (GDP/population).
However, due to the high non-linearity of the map used to calculate countries fitness, we did not expect any trivial correlations to be found.
Indeed, from Figure \ref{fig:Figure6NEW} (a) we can see that country fitness-GDP per capita are only weakly correlated ($R^2=0.194$; Zero correlation test P-value $<0.001$).
As an example, Singapore has high GDP per capita but is an island country with low geographical and climatological diversity, and consequently a small basket of produced food commodities.
On the other hand Bolivia has high fitness (world leader producer of Quinoa), but low GDP per capita.
To further this notion that GDP and fitness are not well related, we can also investigate the dynamic relationship between country fitness and GDP per capita.
Rather than a static representation, we can show country specific trajectories in the plane of country fitness and GDP per capita (three representative countries, China, Brazil and India from 1992 to 2011 shown in Figure \ref{fig:Figure6NEW} (b)).
It is evident that the GDP per capita of these countries increases as time goes on with country specific behavior in terms of fitness.
The fitness of Brazil and India are constant on average, while China's fitness has increased in time along with the increase in GDP per capita, suggesting an improvement in the diversification and relative production volumes (with respect to population) of food commodities.
However, this latter case is an exception rather than a typical behavior: in general a country's fitness is stable in time (not shown here).

\begin{table}[!ht]
\centering
\begin{tabular}{|c|c|c|}
\hline
 & Highest Fitness & Lowest Fitness \\
\hline
1 & Canada & Western Sahara \\
\hline
2 & Bolivia (Plurinational State of) & Djibouti \\
\hline
3 & Australia & Lesotho \\
\hline
4 & New Zealand & Singapore \\
\hline
5 & Poland & Qatar \\
\hline
6 & Peru & Bahrain \\
\hline
7 & Turkey & Puerto Rico \\
\hline
8 & Greece & Equatorial Guinea \\
\hline
9 & Belarus & Comoros \\
\hline
10 & Hungary & Liberia \\
\hline
\end{tabular}
\caption{\label{tab:table2NEW}The 10 highest and lowest fitness countries for the year 2011.}
\end{table}

\begin{table}[!ht]
\centering
\begin{tabular}{|c|c|c|}
\hline
 & Highest Specialization & Lowest Specialization \\
\hline
1 & Quinoa & Sugar cane \\
\hline
2 & Gooseberries & Cassava \\
\hline
3 & Cranberries & Cow milk, whole, fresh \\
\hline
4 & Canary seed & Maize \\
\hline
5 & Blueberries & Rice, paddy \\
\hline
6 & Bird meat, nes & Wheat \\
\hline
7 & Meat of Asses & Potatoes \\
\hline
8 & Brazil nuts, with shell & Bananas \\
\hline
9 & Poppy seed & Vegetables fresh nes \\
\hline
10 & Goose and guinea fowl meat & Plantains \\
\hline
\end{tabular}
\caption{\label{tab:table3NEW}The 10 highest and lowest specialization food for the year 2011.}
\end{table}

We also envision that these quantitative schemes will provide new approaches to fundamental analyses alimented by available data of country food security.
In fact, there are increasing evidences that for a comprehensive analysis of food security and sustainability, food trade must be taken in account, so to evaluate countries dependent on food imports \cite{rosegrant2003global, davis2015accelerated, suweis2015resilience}, and to estimate the impact of intensification of the land grabbing \cite{rulli2013global}.
The narrowing of diversity in crop and animal species contributing to the world's food supplies is considered a potential threat to food security \cite{davis2015livestock, khoury2014increasing} and should be readily identified in the application of the framework above. For example we find that during the period 1992-2011 Ireland and New Zealand decreased their food production diversity, producing 16 and 22 different food commodities in 2011 instead of 21 and 29 in 1991, respectively and with a corresponding $\approx 25\%$ decrease in fitness.
We stress that this analysis can be easily applied to study production and trade networks weighted with respect to different nutritional indicators (proteins, fat, calories).
As such future work can explore import and export data, food specialization with regards to secondary commodities (e.g. chocolates, breads, cheese) and understand how food trade impacts each country's fitness and specialization.
There is also the ability to investigate the relation between food specialization, food price and price volatility as, combined with the presented analysis on import MSF sub-trees, it may help to develop a framework in which to detect early warning signs of local food crises.

\section*{Methods}

\subsection*{Food Production Data and Network Analysis}

Food production and bilateral trade data in tons was extracted from the United Nations Food and Agricultural Organization database (FAOSTAT).
The commodities extracted are comprised only of primary commodities (e.g. wheat) and we do not consider secondary commodities (e.g. bread) in this study.
Fish production data were also extracted from the FAOSTAT database and include the production data for eight primary fish commodities (freshwater fish, crustaceans, demersal fish, pelagic fish, marine fish, cephalopods, mollusks, aquatic mammals).
The data was limited to the interval 1992-2011 and only countries with population greater than half million people were considered.
Country separations and merges during this time period were handled following \cite{carr2013recent} resulting in a final dataset that covers 157 commodities for 177 countries.

From this data the country (c)-food production (p) adjacency matrix ${M_{cp}}\left( y \right)$ is constructed for each year, where ${M_{cp}}\left( y \right)=1$ if country c produces product p, and it zero otherwise. We then build the weighted country-food per capita production matrix $W_{cp}$ where ${W_{cp}}\left( y \right) = \frac{{to{n_{cp}}\left( y \right)}}{{po{p_c}\left( y \right)}}$, where for every year y, $ton_{cp}$ is the production volume (in tons) of the food commodity p produced by country c while, $pop_c$ is the population of country c.
To obtain the country-import graphs ${I_{ci}}\left( y \right)$ we used data on the volumes of food commodities imported by each country for each primary commodity: ${I_{ci}}\left( y \right) = \frac{{to{n_{ci}}\left( y \right)}}{{po{p_c}\left( y \right)}}$.
We finally note that production and import fluxes can be weighted in different ways by appropriately converting the unit of measure (e.g. we can transform tons to calories, by multiplying food commodities volumes by calories content).

The nestedness NODF measure is defined as $NODF = \frac{{\sum\nolimits_{i < j:i,j \in C} {T_{ij}^C}  + \sum\nolimits_{i < j:i,j \in P} {T_{ij}^P} }}{{\left[ {\frac{{C\left( {C - 1} \right)}}{2}} \right] + \left[ {\frac{{P\left( {P - 1} \right)}}{2}} \right]}}$ \cite{almeida2008consistent}, where $T_{ij}^X = 0$ if $k_i^X = k_j^X$  and $T_{ij}^X = {{o_{ij}^X} \mathord{\left/ {\vphantom {{o_{ij}^X} {\min \left( {k_i^X,k_j^X} \right)}}} \right. \kern-\nulldelimiterspace} {\min \left( {k_i^X,k_j^X} \right)}}$ when both i and j belong to the same set $X=C,P$ (the number of countries and food products respectively).
To understand if the country - food production graph is nested, we first calculate the NODF of  , we then build 100 random networks (where links are placed at random) with the same size and connectance, and calculate the NODF of these graphs.
We finally compare the NODF of empirical data versus the distribution of NODF values corresponding to the null models.
As discussed in the Results section, we find that the food-production network is significantly nested, and the countries that produce high specialized food products have a tendency to produce also low specialized food products.

\subsection*{Minimum Spanning Forest Algorithm}

From the bipartite food-production graph, we can build the corresponding projection graphs \cite{newman2010networks}. The country-country network is characterized by the ${N_C} \times {N_C}$ country-country matrix $C = W{W^T}$.
The non-diagonal elements $C_{cc'}$ correspond to the number of products that countries $c$ and $c'$ have in common.
The diagonal elements $C_{cc}$ corresponds to the number of products produced by country c and are a measure of the diversification of country c.
To quantify the correlation in the production among two countries, we can define the similarity matrix among countries as   ${S_{cc'}} = \frac{{2{C_{cc'}}}}{{{C_{cc}} + {C_{c'c'}}}}$, where $0 \le {S_{cc'}} \le 1$ and the larger the value, the stronger is the correlation between the food commodities produced by the two countries $c$ and $c'$.
In particular, from the correlation matrix S, using the method known as minimum spanning tree \cite{caldarelli2012network}, it is possible to divide the country-country networks in different ?communities characterized by a strong overlap in the food production patterns.
The same analysis can be performed to the food-food network: in this case the communities represent baskets of food products that share very similar producers. The algorithm is structured performing the following steps:

\begin{enumerate}
  \item Sorting all the edges of the similarity matrix S by their weight (from the largest weight to the smallest one)
  \item Adding one edge at each time from the sorted vector obtained in Step 1. If the two nodes of the new edge have been matched by the previous edges, that is to say the two nodes belongs to the previous edges, removing this edge
  \item Repeating Step 2 until all the edges are added
\end{enumerate}

\subsection*{Calculation of the fitness of countries and the specialization of food products}
We here describe the non-linear iterative procedure proposed in Tacchella et al.\cite{tacchella2012new} that we then apply to determine the fitness of countries and the specialization of food products.
In mathematical terms, fitness and specialization define a new metric for determining the relative centrality of countries and products in the context of food production.
This non-linear relation can be seen as the fixed-point equation of an iterative algorithm so that the quantities $Q_p$ and $F_c$ are quantitatively estimated through the attractive asymptotic fixed point of a system of coupled equation.
Moreover, each iteration of the algorithm adds higher order information on these quantities, eventually reaching fat tail log-Normal distributions for the two metrics at the fixed point. The iteration process results in a unique set of two fixed points for any initial condition.
The algorithm to determine food specialization and country fitness \cite{tacchella2012new} can be summarized in the following steps:

\begin{enumerate}
  \item Setting the initial condition of country fitness $Q_p^0 = 1$ and production specialization $F_c^0 = 1$
  \item Computing the intermediate variables as follows $\left\{ {\begin{array}{*{20}{c}}
{\tilde F_c^n = \sum\nolimits_p {{W_{cp}}Q_p^{ - 1}} }\\
{\tilde Q_p^n = \frac{1}{{\sum\nolimits_c {{W_{cp}}\frac{1}{{F_c^{n - 1}}}} }}}
\end{array}} \right.$, where $W_{cp}$ is the element of the weighted country-food product matrix W, $\tilde F_c^n$ and $\tilde Q_p^n$ are the computing intermediate variables, n is the iterative step. For better parallelization computing, this step could also change to vector calculation as follows $\left\{ {\begin{array}{*{20}{c}}
{{{\bf{F}}^n} = {\bf{W}}{{\bf{Q}}^{n - 1}}}\\
{{{\bf{Q}}^n} = \frac{1}{{{{\bf{W}}^T}\frac{1}{{{{\bf{F}}^{n - 1}}}}}}}
\end{array}} \right.$
  \item Normalizing the intermediate variables of country fitness and production specialization $\left\{ {\begin{array}{*{20}{c}}
{F_c^n = \frac{{\tilde F_c^n}}{{\left\langle {\tilde F_c^n} \right\rangle }}}\\
{Q_p^n = \frac{{\tilde Q_p^n}}{{\left\langle {\tilde Q_p^n} \right\rangle }}}
\end{array}} \right.$
  \item Repeating until the country fitness $F_c^n$ and production specialization $Q_p^n$ are fixed
\end{enumerate}

\section*{Acknowledgements}

S.S. acknowledges P. D'Odorico and A. Maritan for useful discussions.

\bibliography{Tu_et_al_plos2016}


\end{document}